%% file: main.tex
\documentclass[conference]{IEEEtran}
\IEEEoverridecommandlockouts

\usepackage{cite}
\usepackage{amsmath,amssymb,amsfonts}
\usepackage{bbm}
\usepackage{algorithm}
\usepackage[noEnd,indLines]{algpseudocodex}

\renewcommand{\baselinestretch}{0.98}

\newtheorem{Theorem}{Theorem}

\newtheorem{Lemma}{Lemma}

\allowdisplaybreaks[4]

\usepackage{booktabs} 
\usepackage{multirow}
\usepackage{graphicx}
\usepackage{textcomp}
\usepackage{xcolor}
\def\BibTeX{{\rm B\kern-.05em{\sc i\kern-.025em b}\kern-.08em
    T\kern-.1667em\lower.7ex\hbox{E}\kern-.125emX}}
\begin{document}

\title{Network Diffuser for Placing-Scheduling Service Function Chains with Inverse Demonstration \\
}

\author{\IEEEauthorblockN{1\textsuperscript{st} Zuyuan Zhang}
\IEEEauthorblockA{\textit{The George Washington University} \\
zuyuan.zhang@gwu.edu}
\and
\IEEEauthorblockN{2\textsuperscript{nd} Vaneet Aggarwal}
\IEEEauthorblockA{\textit{Purdue University} \\
vaneet@purdue.edu}
\and
\IEEEauthorblockN{3\textsuperscript{rd} Tian Lan}
\IEEEauthorblockA{\textit{The George Washington University} \\
tlan@gwu.edu}
}

\maketitle

\input{01abstract}

\input{02introduction}

\input{03background}

\input{04model}

\input{05algorithm}

\input{06simulation}

\bibliographystyle{IEEEtran}
\bibliography{main}

\vspace{12pt}

\end{document}

%% file: 01abstract.tex
\begin{abstract}

Network services are increasingly managed by considering chained-up virtual network functions and relevant traffic flows, known as the Service Function Chains (SFCs). To deal with sequential arrivals of SFCs in an online fashion, we must consider two closely-coupled problems – an SFC placement problem that maps SFCs to servers/links in the network and an SFC scheduling problem that determines when each SFC is executed. Solving the whole SFC problem targeting these two optimizations jointly is extremely challenging. In this paper, we propose a novel network diffuser using conditional generative modeling for this SFC placing-scheduling optimization. Recent advances in generative AI and diffusion models have made it possible to generate high-quality images/videos and decision trajectories from language description.  We formulate the SFC optimization as a problem of generating a state sequence for planning and perform graph diffusion on the state trajectories to enable extraction of SFC decisions, with SFC optimization constraints and objectives as conditions. To address the lack of demonstration data due to NP-hardness and exponential problem space of the SFC optimization, we also propose a novel and somewhat maverick approach -- Rather than solving instances of this difficult optimization, we start with randomly-generated solutions as input, and then determine appropriate SFC optimization problems that render these solutions feasible. This inverse demonstration enables us to obtain sufficient expert demonstrations, i.e., problem-solution pairs, through further optimization. In our numerical evaluations, the proposed network diffuser outperforms learning and heuristic baselines, by $\sim$20\% improvement in SFC reward and $\sim$50\% reduction in SFC waiting time and blocking rate.

\end{abstract}


%% file: 02introduction.tex
\section{Introduction}

To enable a vision toward future networks with full autonomy, the focus of network service management has been quickly shifting toward Network Function Virtualization (NFV)~\cite{cziva2018dynamic,fei2018adaptive,shin2020inferring}, which enables migration of network functions and middle-boxes to generic-purpose servers located in cloud nodes~\cite{DBLP:conf/infocom/TomassilliGHP18}. By chaining up virtual network functions (e.g., firewall, caching, content filtering, and inference) as ordered sequences and supporting the traffic flows between adjacent functions, the notion of Service Function Chains (SFCs) makes network services more cost-efficient, elastic, and flexible. But it also gives rise to a challenging optimization problem -- how to map SFCs to servers and links in the network, to efficient utilize available computing and network resources while achieving specific design objectives. 

Many SFC optimization problems are known to be NP-hard. Existing work~\cite{pei2018efficiently,ren2020efficient, wang2019reinforcement,DBLP:journals/tpds/YuZSMSHBWW21,DBLP:conf/infocom/MaoSY22a,DBLP:conf/infocom/TomassilliGHP18} often rely on approximation algorithms to find solutions that can be within a certain range of performance for various special cases. Examples include heuristic methods~\cite{pei2018efficiently,ren2020efficient} for minimizing SFC resource utilization/congestion and dynamic packing algorithms~\cite{DBLP:conf/infocom/MaoSY22a} achieving 2-approximation solutions in SFC placement and flow routing. It is shown in \cite{DBLP:conf/infocom/TomassilliGHP18} that the general SFC placement problem can be modeled as a Set Cover Problem~\cite{DBLP:conf/conext/ChaudetFLRV05} and thus cannot achieve a better-than-logarithmic approximation factor. However, to deal with sequential arrivals of SFCs in
an online manner, we must consider two closely-coupled problems -- an {\bf SFC placement problem} that maps SFCs to servers/links in the network and an {\bf SFC scheduling problem} that determines when each SFC is executed. Solving the whole SFC problem targeting these two optimizations jointly is extremely challenging. Most existing work are limited to one individual optimization. 

Machine Learning (ML) is a natural approach to this difficult SFC placing-scheduling optimization. For many network optimization problems such as routing and load-balancing, learning algorithms, e.g., \cite{fadlullah2017state, watkins1992q, slivkins2019introduction,he2023routing,almasan2022enero,jiang2019reinforcement,feng2009reinforcement,zhang2024collaborative,zhang2024modeling,qiao2024br,zou2024distributed,jing2024byzantine,zhang2024distributed,mei2024projection}, have demonstrated tremendous promise for making it possible to turn large training datasets into powerful decision making policies. However, the exponentially-large problem space and NP-hardness of SFC optimization makes it challenging to obtain expert demonstration data (i.e., problem-solution pairs) for training -- using either heuristic methods or exploration policies like reinforcement learning~\cite{wang2019reinforcement}. This creates a chicken-and-egg problem, i.e., training a learning
model for SFC optimization requires having sufficient expert
demonstration, which however cannot be obtained without
having a good solution first.

In this paper, we consider the SFC placing-scheduling optimization and propose a novel {\bf network diffuser} using conditional generative modeling for SFC optimization. Recent advances in generative AI and diffusion models have made it possible to generate high-quality images/videos and decision trajectories from language descriptions~\cite{DBLP:journals/corr/abs-2211-15657}. We formulate an SFC optimization as the problem of generating a state sequence $x(t) = (s_{t},s_{t+1},\dots,s_{t+H-1})$ from current time $t$ into the $H$-step future. Each state $s_t$ encapsulates both network state (e.g., topology, server resource and link utilization) and SFC status (e.g., requirements, placement, and schedules). Thus, by incorporating SFC optimization constraints and objectives as conditions for guided diffusion~\cite{ho2020denoising}, we can perform diffusion on the state trajectories and learn a generative model of the distribution $q(x)$ from the dataset $\mathcal{D} = \{x_{i}\}_{0\leq i \leq M}$ that consists of available expert demonstrations (i.e., SFC problem-solution pairs). This allows us to generate $x(t)$ containing future states of the SFC optimization and then extract SFC placing-scheduling decisions from the generated states. Further, to efficiently represent network information in our problem, we leverage graph diffusion with a discrete-denoising diffusion model~\cite{vignac2022digress} for more efficient embedding of the SFC optimization on networks. To the best of our knowledge, this is the first proposal of a network diffuser for the SFC placing-scheduling optimization.

To address the lack of demonstration due to NP-hardness and exponential problem space of SFC optimization, we propose a novel and somewhat maverick approach for generating expert demonstrations. Rather than solving instances of the difficult SFC optimization, our idea, denoted as {\bf inverse demonstration}, reverses this process by starting with (randomly generated) SFC placement/scheduling decisions as input, and then determines an appropriate SFC optimization problem that render these decisions feasible. This way, we can easily obtain a problem-solution pair of the SFC optimization in the reverse direction. We further optimize the demonstration through a lexcographic min-max optimization of the SFC completion times, which is shown to be transformed into an integer programming with separable convex objective functions. This novel inverse demonstration approach enables us to obtain sufficient expert demonstrations of the SFC optimization. It is applicable to many networking problems where demonstrations are difficult to obtain.

The proposed network diffuser is evaluated in a simulated environment and compared with several heuristic and deep learning baselines, such as~\cite{yuan2023deploy} for varying network size and number of SFC requests. In our numerical evaluations, the proposed network diffuser outperforms the baselines, by $\sim$20\% improvement in SFC reward and 50\% reduction in both SFC waiting time and blacking rate. The benefits of inverse demonstration for generating expert training data is also validated showing $\sim$15\% improvement in reward.
The key contributions of our paper are summarized as follows:
\begin{itemize}
    \item We solve an SFC optimization that jointly targets two closely-coupled problems: SFC placement and SFC scheduling, with respect to sequential arrivals of SFCs in an online manner with non-zero release times.
    \item We propose a novel network diffuser using conditional generative modeling for the SFC placing-scheduling optimization. It leverages graph diffusion to efficiently embed network information and performs diffusion on the state trajectories for extracting decisions. 
    \item To obtain demonstration data for training without needing to solve the difficult optimization, we propose a somewhat maverick approach that starts with randomly generated decisions and then finds an appropriate SFC optimization problem, denoted as inverse demonstration.
   \item We implement the proposed network diffuser and compare it with several heuristic and deep learning baselines. Significant performance improvements are observed.
\end{itemize}

%% file: 03background.tex
\section{Background and Related Work}

\noindent {\bf SFC Placement and Scheduling.}
The objective of the SFC placement and scheduling problem is to (i) optimally map the network functions and associated data flows of a set of SFCs to the server nodes and links of the underlying physical network and (ii) schedule the SFC executions over time, with the goal of minimizing congestion, resource cost, and blocking rate. This problem is becoming increasingly important to fully capitalize on the flexible and elasticity of SFC-based network management.
However, many SFC placement-scheduling problems are NP-hard and lack efficient solutions.
Existing work often rely on approximation algorithms to find solutions that can be proven to be within a certain range of quality and performance, e.g.,~\cite{DBLP:journals/tpds/YuZSMSHBWW21,DBLP:conf/infocom/MaoSY22a,DBLP:conf/infocom/TomassilliGHP18,mao2022joint,blocher2020letting,jaisudthi2023profiler}. In particular, the problem of online joint SFC placement and flow routing is considered in~\cite{DBLP:conf/infocom/MaoSY22a,mao2022provably,cao2023sfc,sallam2018shortest} with a two-stage algorithm achieving an approximation ratio of two. The authors in~\cite{DBLP:conf/infocom/TomassilliGHP18} show that SFC placement under order constraints can be seen as an instance of the Set Cover Problem~\cite{DBLP:conf/conext/ChaudetFLRV05}. Thus, an algorithm cannot achieve a better approximation factor than $(1-\epsilon)\ln{|V|}$ (where $|V|$ is the network size) unless $P=NP$, even if all the SFCs consist of only one function. Learning-based solutions, such as deep reinforcement learning~\cite{wang2019reinforcement,yuan2023deploy}, have also been considered. While these learning solutions have demonstrated great promise, they face several key challenges, including limited data and insufficient exploration. In this paper, we propose a novel approach using decision diffusion -- a form of generative AI technique -- for SFC placement and scheduling.

\noindent {\bf Learning-based network management.} 
Deep learning (Dl) models can effectively analyze large amounts of network data, identifying patterns and anomalies~\cite{wang2021machine,zhou2024real,chen2024survey}. Additionally, DL has been used for network configuration optimization, enabling automated and adaptive network management~\cite{fadlullah2017state}.
Reinforcement learning plays a key role in achieving adaptive and real-time control of network parameters. Techniques such as Q-learning~\cite{watkins1992q}, multi-armed bandit (MAB)~\cite{slivkins2019introduction} models, and actor-critic~\cite{konda1999actor} methods have been used to optimize routing~\cite{he2023routing,almasan2022enero}, load balancing~\cite{jiang2019reinforcement,zhang2024distributed}, and fault tolerance in wired networks~\cite{feng2009reinforcement,ravari2024adversarial}. These methods enable the network to adapt to different conditions and improve overall performance. Network topology in these approaches is often represented as a graph. 
However, online learning techniques often struggle with exploration of large state/decision space of SFC placing-scheduling, while offline learning faces the lack of expert demonstration data, due to the NP-hardness of the problem.

\noindent {\bf Decision Diffuser.}
Recent advances in generative AI -- e.g., diffusion models --- have made it possible to generate high-quality images/videos from language descriptions. It is shown that these methods can directly address the problem of sequential decision-making in many challenging domains~\cite{DBLP:journals/corr/abs-2211-15657,yu2025look}, outperforming existing approaches like offline reinforcement learning. The diffusion model\cite{ho2020denoising} is a specific type of generative model that learns the data distribution $q(x)$ from the dataset $\mathcal{D} = \{x_{i}\}_{0\leq i \leq M}$.
The data generation process uses a predefined forward noise process $q(x_{k+1}|x_{k}) = \mathcal{N}(x_{k+1};\sqrt{\alpha_{k}}x_{k},(1-\alpha)I)$ and a trainable reverse process $p_{\theta}(x_{k-1}|x_{k})=\mathcal{N}(x_{k-1}|\mu_{\theta}(x_{k},k),\Sigma_{k})$, where $\mathcal{N}(\mu,\Sigma)$ represents a normal distribution with mean $\mu$ and covariance $\Sigma$, $\alpha_{k}\in \mathbb{R}$ determines the variance schedule.
To generate the desired sample $x_{0}=x$, we sample $x_{K}\sim\mathcal{N}(0,I)$. Then, through a denoising process, we create a series of intermediate samples $x_1,x_2,...,x_{K-1}$, which leads to the generation of desired sample. To the best of our knowledge, this is the first work proposing a diffusion-modeling approach to the SFC placement and scheduling problem.

\noindent {\bf Guided Diffusion and Graph Diffusion.}
The equivalence between diffusion models and score matching~\cite{song2020denoising} indicates $\epsilon_{\theta}(x_{k},k)\propto \nabla_{x_{k}}\log p(x_{k})$, leading to two conditioning methods: classifier guidance ~\cite{nichol2021improved} and classifier-free guidance ~\cite{ho2022classifier}.
The former requires training an additional classifier $p_{\phi}(y|x_{k})$ on the noisy data.
The latter does not train a classifier separately but modifies the original training setup to learn both the conditional $\epsilon_{\theta}(x_{k},y,k)$ and unconditional $\epsilon_{\theta}(x_{k},\varnothing,k)$ noise models.
The perturbed noise $\epsilon_{\theta}(x_{k},k)+w(\epsilon_{\theta}(x_{k},y,k)-\epsilon_{\theta}(x_{k},k))$ is used for sample generation later.
Previous graph diffusion models suggest embedding the graph into a continuous space, then performing a similar diffusion process to that of images, and adding Gaussian noise to node features and the graph adjacency matrix ~\cite{niu2020permutation}.
A discrete denoising diffusion model is employed in~\cite{vignac2022digress} to generate graphs with categorical node and edge attributes. Training diffusion models often requires large amount of high-quality data. This is a bottleneck for SFC placing-scheduling, where limited expert demonstrations are available to its NP-hardness.

%% file: 04model.tex
\section{MODEL AND PROBLEM DEFINITION}
The problem of placing and scheduling a set of SFCs on a network is reprented by a tuple $(G,F,T)$, including a network $G$, a set of SFCs $F$, and a deadline $T$. The network is represented by an undirected graph $G = (V;E)$, where $V$ is a set of nodes, each representing a server that can host a range of network functions (e.g., caching, firewall, encryption, computation, inference) with different resource requirements, 
and $E$ is a set of network links connecting nodes in $V$ with different bandwidth available. 
We consider heterogeneous networks -- in which each server $j\in V$ has server resource  $C_j^V$ (which could become a vector if multiple resources~\cite{joe2013multiresource} are considered, e.g., storage and computation), while for any $(p, q) \in E$, there exist a communication link between servers $p$ and $q$, with a bandwidth constraint of $B_{p,q}^E$ on link $(p,q)$. For any $(p, q) \notin E$, we simply denote the available bandwidth as zero.

We consider a set of $M$ SFCs, denoted by $F=\{1,2,\ldots,M\}$. Each SFC $i\in F$ has a duration of $D_i$ seconds and a weight $w_i$, and can start execution after arriving at time $t_i^s$, which is known as the the release time in scheduling problems~\cite{aggarwal2021preemptive,aggarwal2021approximability}. It contains a sequence of virtual network functions (VNFs) -- e.g., firewall, caching, computation, and inference -- chained in some order, with corresponding network traffic flow between the functions. Let $L_i$ be the length of SFC $i$, i.e., containing $L_i$ nodes representing the VNFs and $L_i-1$ edges representing the corresponding traffic flows. We use $c_{i,j}$ to denote the resource requirement of the $j$th node of SFC $i$ and $b_{i,j}$ to denote the bandwidth requirement of the $j$th flow of SFC $j$. We note that resource requirement $c_{i,j}$ can be a vector if multiple resources are considered. In our problem formulation, the required bandwidth $b_{i,j}$ can change along the SFC (for different flow $j$), e.g., due to network functions such as up-sampling or rendering (which increases the output data rate) and feature-extraction or compression (which decreases the output data rate).

Our goal is to maximize the number of SFCs that can be successfully processed on a given network within deadline $T$. An SFC 
can be successfully placed and scheduled if we find a feasible solution that satisfies both server resource and bandwidth constraints, $C_j^V$ and $B_{p,q}^E$. For SFC placement, we need to determine the mapping of SFCs (i.e., their nodes and flows) to the network $G$. For SFC scheduling, we need to find the resource allocation over time to satisfy duration $D_i$ and release time $t_i^s$. To make it more rigorous, we model this as an integer optimization with respect to placement variables $z_{i,j}^p\in\{0,1\}$ and scheduling variables $x_{i,t}\in\{0,1\}$, which are dependent and jointly define our decision making. More precisely, we have $x_{i,t}=1$ if SFC-$i$ is scheduled to be active at time $t$ and $x_{i,t}=0$ otherwise. Further, we have $z_{i,j}^p=1$ if node $j$ of SFC-$i$ is placed on server $p\in V$ of the network and $z_{i,j}^p=0$ otherwise. It is easy to see that  the product $x_{i,t} z^p_{i,j}\in\{0,1\}$ provides an indicator of whether server $p$ is occupied by SFC-$i$ at time $t$. Similarly, $x_{i,t} z^p_{i,j}z^q_{i,j+1}\in\{0,1\}$ indicates whether a link $(p,q)$ on the network is occupied by the $j$th flow of SFC-$i$ (i.e., when servers $p$ and $q$ are occupied by two adjacent nodes, node $j$ and node $j+1$, of the SFC). 

These result in the SFC optimization problem as follows:
\begin{eqnarray}
    & \max & \sum\limits_{i}^{M} I_{i} \label{opt1} \\
    & & \sum_{i,j} c_{i,j} x_{i,t} z^p_{i,j} \le C_p^V, 
 \ \forall p,t \label{opt1_1} \\
    & & \sum_{i,j} b_{i,j} x_{i,t} z^p_{i,j} z^q_{i,j+1} \le B_{p,q}^E, \ \forall (p,q),t \label{opt1_2} \\
    & & \sum_{t=t_i^s}^T x_{i,t} = D_i, \ \forall i,j,p \label{opt1_3} \\
    & & \sum_{p} z^p_{i,j} \le  1, \ \forall i,j \label{opt1_4} \\
    & & I_i = \prod_j \left(\sum_{p} z^p_{i,j}\right), \ \forall i \label{opt1_5} \\
    & & I_i,x_{i,t},z^p_{i,j} \in \{0,1\}, \ \forall i,j,p,t \label{opt1_6}
\end{eqnarray}
were inequalities (\ref{opt1_1}) and (\ref{opt1_2}) represent the server resource and bandwidth constraints, respectively; Constraint (\ref{opt1_3}) ensures that enough execution time of $D_i$ is assigned to SFC-$i$; and Constraint (\ref{opt1_4}) requires each node of the SFC is mapped to at most 1 server. $I_i = \prod_j \left(\sum_{p} z^p_{i,j}\right) \in \{0,1\}$ is an integer variable, and $I_i=1$ only if SFC-$i$ is successfully placed -- when each node $j$ of the SFC is mapped to some server (i.e., $\sum_{p} z^p_{i,j} =1$ for all $j$ along this SFC), as defined in (\ref{opt1_5}). The goal of this SFC optimization is to maximize the number of SFCs successfully placed and scheduled before a given deadline $T$. We note that in our fomulation, it is possible to place adjacent nodes $j$ and $j+1$ of SFC-$i$ on the same server. In this case, the data flow between them is not transmitted on any links, so that it will not cause a traffic burden on the physic network. We call such data flows the idle data flows. Further, since $G$ may not be a complete graph, placing a flow on link $(p,q)$ is possible only if there is a edge between the two servers in $G$, with non-zero bandwidth $B_{p,q}^E$.

\vspace{-0.2in}
\begin{figure}[th] 
\centering 
\includegraphics[width=0.9\linewidth]{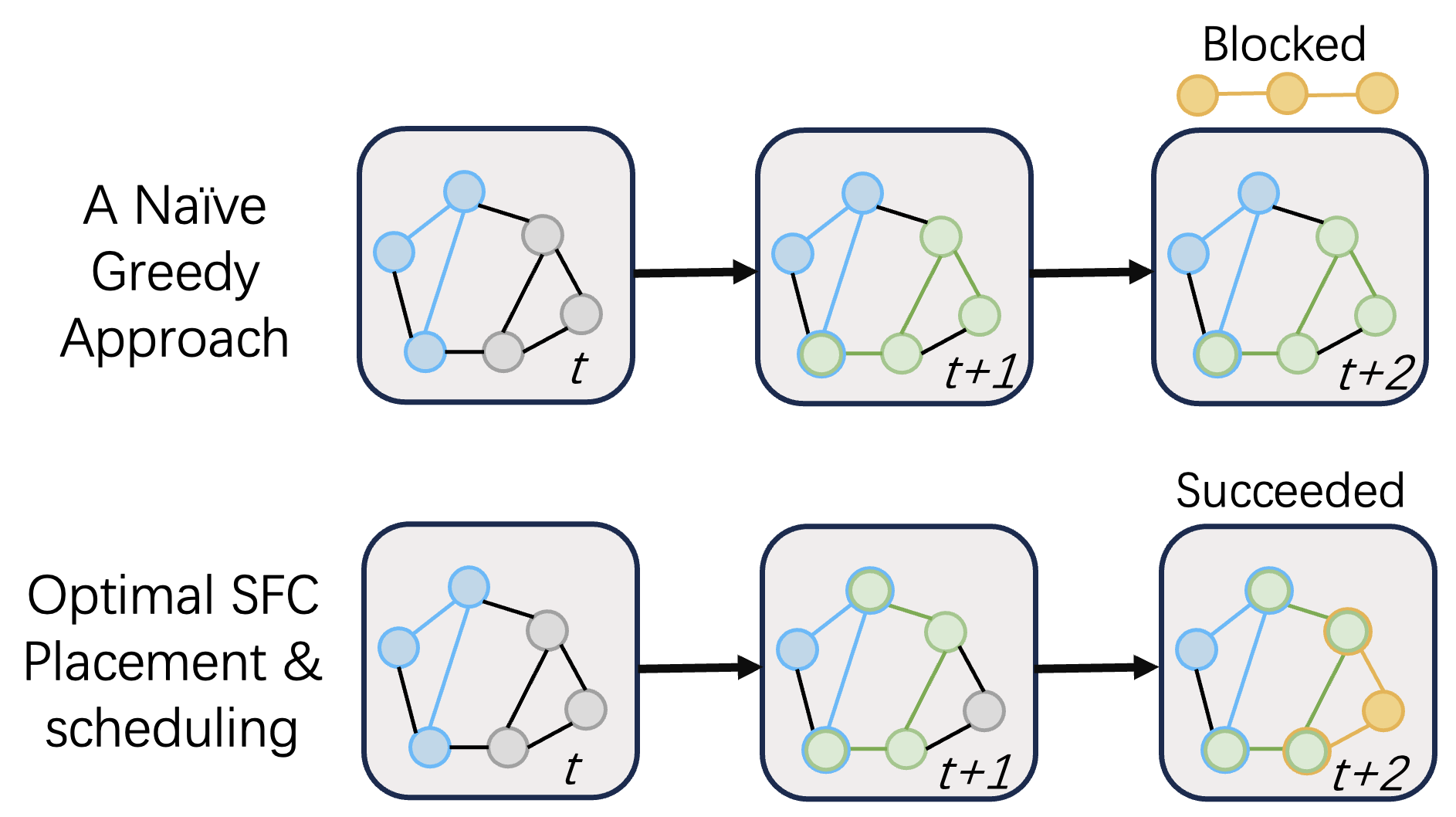} 
\vspace{-0.1in}
\caption{An illustration of the SFC placing-scheduling optimization problem for a toy network of 6 nodes and 8 links. Three SFCs, denoted by Blue, Green, and Yellow, arrive sequntially, one at each time step. A naive greedy approach will result in SFC Yellow being blocked, while a solution jointly optimizing both SFC placement and scheduling can accomadate all three SFCs. We propose a network diffuser to solve it using generative models, by generating a sequence of system states as shown in this figure with constraint-guided diffusion and then extracting optimal placing-scheduling decisions from them.
} 
\label{fig:example} 
\end{figure}
\vspace{-0.05in}

Figure~\ref{fig:example} shows a toy example of an SFC placing-scheduling problem on a network with 6 nodes and 8 links.
Suppose that each node has a computing resource constraint of 2 and each link has a bandwidth constraint of 1. There are three SFCs arriving at the network at time $t$, $t+1$, and $t+2$, respectively, which are denoted by Blue, Green, and Yellow. SFCs Blue and Yellow both have 3 nodes and 2 flows connecting them, with unit computing and bandwidth requirements, while SFC Green has 4 nodes and 3 flows also with with unit computing and bandwidth requirements. It is easy to see that a naive greedy approach trying to minimize resource congestion would cause SFC Yellow to be blocked at time $t+2$. In contrast, an optimal solution to this SFC placing-scheduling problem can accommodate all 3 SFCs. The placement problem and scheduling problem are closely-coupled over time and must be solved jointly to obtain optimal performance. Next, we will show that this SFC placing-scheduling problem is indeed NP-hard and propose a network diffuser to solve it using generative models. It generates a sequence of system states 
with constraint-guided diffusion, so that optimal placing-scheduling decisions can be extracted from the generated states. We will use the network in Figure~\ref{fig:example} as a running example to later illustrate our network diffusor representation and our evaluation results.

\section{Our Solution}
\begin{figure*}[th] 
\centering 
\includegraphics[width=0.85\linewidth]{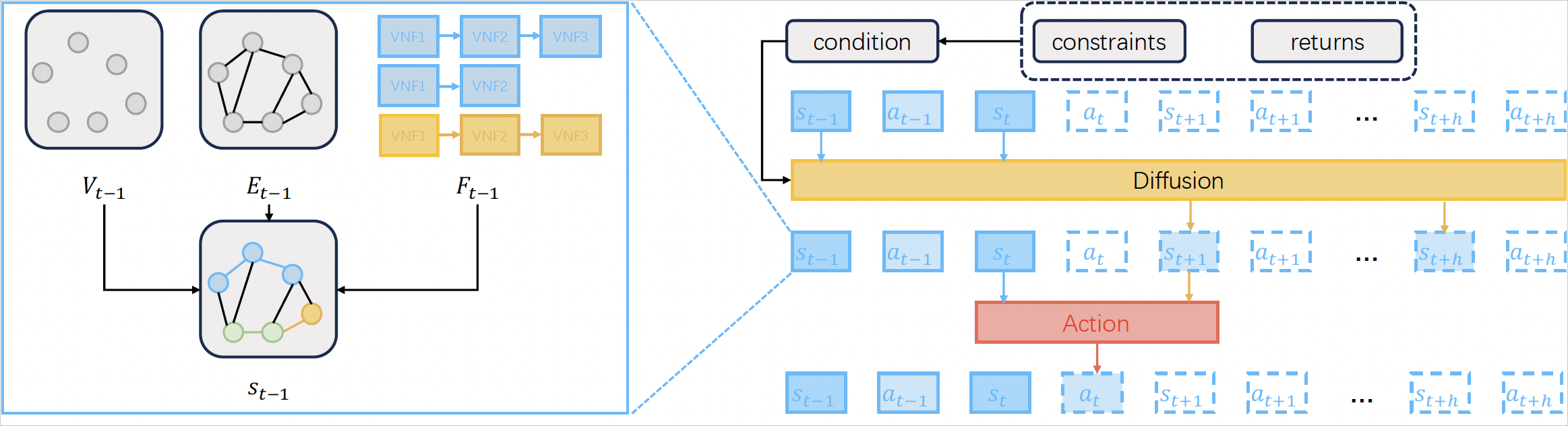} 
\caption{An illustration of our proposed network diffuser for SFC optimization. Each state $s_{t}$ in the trajectory encapsulates server state $V_{t}$ (such as server utilization), link state $E_{t}$ (such as link utilization and network topology), and SFC information $F_{t}$ (such as SFC parameters, placement, and schedules), with respect to the SFC optimization variables $x_{i,t}$ and $z_{i,j}^p$. The SFC optimization constraints and objective/reward value are represented by conditions as guidance/input to the diffusion model. Given the current state $s_t$ and conditioning, our network diffuser generates a sequence of future system states. It then extracts and executes the SFC placement and scheduling actions (i.e., $x_{i,t}$ and $z_{i,j}^p$) at that leads to the immediate future state $s_{t+1}$. } 
\label{fig:algorithm1} 
\vspace{-0.5cm}
\end{figure*}

We propose a novel framework that leverages diffusion models -- a generative AI technique -- to solve the proposed SFC optimization. To this end, we formulate the sequence of states (denoted by $s_t=[G_t,F_t]$ capturing the network and SFC states at $t$) of the SFC optimization over time $t$ as a trajectory. In images, the diffusion process is applied across all pixel values in an image. It is therefore  natural to apply a similar diffusion process on the SFC optimization trajectory. As shown in Figure~\ref{fig:algorithm1}, we develop a network diffuser using graph generation algorithms to capture both network and SFC representation and incorporate SFC optimization constraints/parameters as conditions for guided diffusion. By performing diffusion on the state sequence, we can generate future states of the SFC optimization and then extract SFC decisions/actions from the generated states. This approach enables highly efficient solutions for the SFC optimization problems, under given constraints, network topolgies, and SFC requirements, which is shown to be NP-hard.

\vspace{0.05in}
\begin{Theorem}
\label{the:NP-hard}
The proposed SFC Optimization for placement and scheduling in (\ref{opt1})-(\ref{opt1_6}) is NP-hard.
\end{Theorem}
\begin{IEEEproof}
We show that a multi-dimensional knapsack problem can be cast into the proposed SFC optimization. Let $c_{i,j}$ represent the size of item $i$ over dimension $j$, $C_p^V$ be the capacity of knapsack $p$, and $w_i$ be the value of item $i$. For sufficiently large $B_{p,q}^E$ and $D_i=T$ for all SFCs, it is easy to see that $I_i$ represents whether item $i$ is selected and $\sum_i I_i$ denotes the total number of items in the knapsack (i.e, subset-sum problem). Thus, we can convert the multi-dimensional knapsack problem into the SFC optimization in polynomial time. Thus the SFC optimization is NP-hard.
\end{IEEEproof}
\vspace{0.05in}

While diffusion models have achieved state-of-the-art performance on image/video generation tasks, they require having access to huge volumes of training data~\cite{DBLP:journals/corr/abs-2211-15657}. This poses a serious challenge to using diffusion models for the proposed SFC placement and scheduling problem. Specifically, there exists a chicken-and-egg problem, i.e., training a diffusion model for SFC optimization requires having sufficient expert demonstration, which however cannot be obtained without having an efficient solution first. Since the proposed SFC optimization is NP-hard and has an exponential problem space of size $|V|^{LM}\cdot T^M$, where $|V|$ is the size of the network, $L$ length of SFC, $M$ the number of SFCs, and $T$ the length of time horizon, obtaining expert demonstrations for training (either using heuristics or reinforcement learning based exploration) is indeed difficult.

To address this challenge, we propose a novel approach for generating expert demonstrations in SFC optimization. Our approach, denoted as inverse demonstration, is inspired by the idea of inverse optimization~\cite{chan2023inverse}.
As shown in Figure~\ref{fig:algorithm2}, rather than seeking to computing a given SFC optimization for obtaining demonstrations -- which is NP-hard and has exponential problem space of size $|V|^{LM}\cdot T^M$, we take (randomly generated) SFC placement/scheduling decisions as input and determines SFC optimization problems (i.e., with parameters $(G,F,T)$) that render these decisions nearly optimal (using a lexicographic min-max optimization of the SFC completion times). It leads to problem-solution pairs (in the reverse direction) with expert level performance, which are leveraged to train our diffusion models.

%% file: 05algorithm.tex
\subsection{Network Diffuser for SFC Optimization}

To solve the SFC deployment problem, we propose the SFC Decision Generation Algorithm.
This is a Decision Diffusion algorithm based on graph generation. 
Next, we discuss how to use diffusion for decision-making.
First, we discuss the modeling choices for diffusion.
Next, we will discuss how to capture the best aspects of the trajectory using classifier-free guidance.
Then, we discuss the different behaviors that can be achieved using conditional diffusion models.
Finally, we discuss the practical training details of our approach.

We use $q$ to represent the forward denoising process, and $p_{\theta}$ to represent the reverse denoising process.

\subsubsection{Diffusing over states}

\begin{figure}[th] 
\centering 
\includegraphics[width=0.95\linewidth]{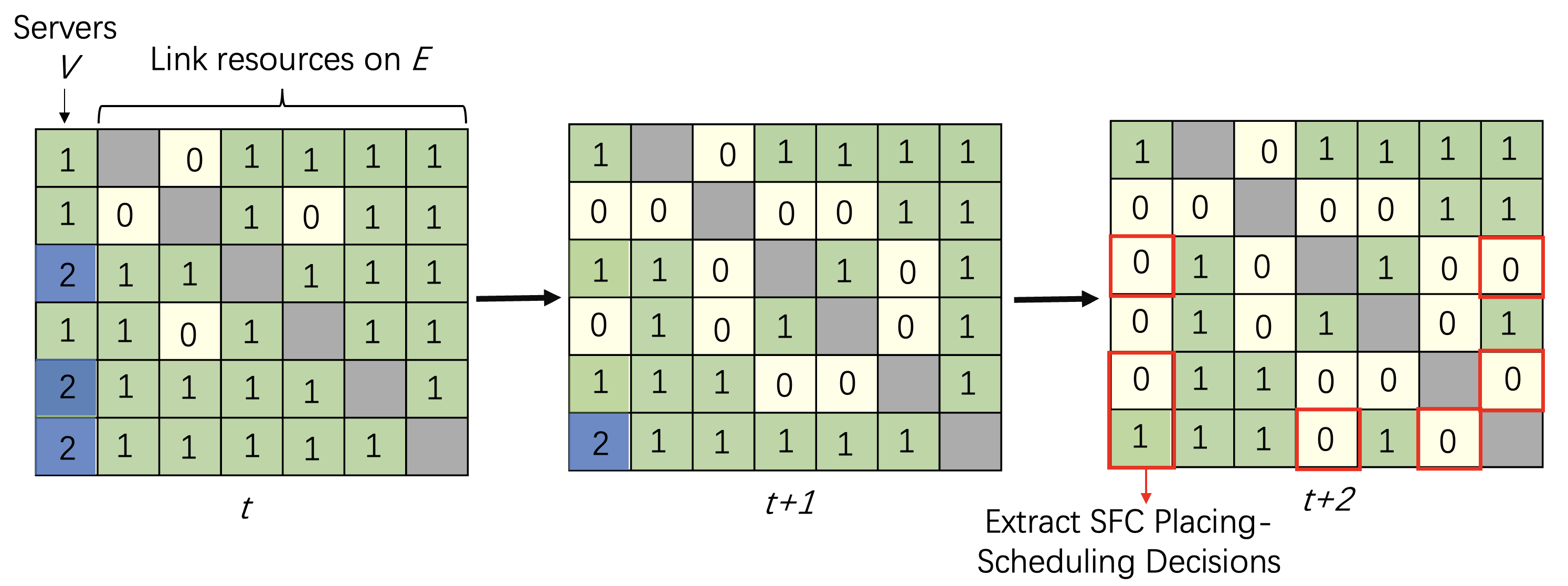} 
\vspace{-0.15in}
\caption{A visualization of our state sequence representation in our network diffuser, for the optimal SFC placing-scheduling solution shown previously in Figure~\ref{fig:example2}. We will perform diffusion over these state trajectories, to generate future states and extract SFC placing-scheduling decisions from them.
} 
\label{fig:example2} 
\end{figure}

The diffusion process requires a natural simulation of trajectory and actions. In the SFC environment, the state often does not change drastically. However, the actions that determine state changes are often diverse and not very smooth, making them more difficult to predict and model. Therefore, we only perform diffusion on the state. 
Here we consider the placing-scheduling variables $z_{i,j}^p$ and $x_{i,t}$ in our SFC optimization, i.e., (\ref{opt1})-(\ref{opt1_6}), and embed them into state (and trajectory) representations. More precisely, we divide the state representation of the SFC into two parts. The first part is the information of the graph, which includes the connectivity information and the current network state including remaining resources on the servers and links. The second part is the SFC parameter/configuration, which includes the SFC itself, the starting node, and the resource requirements of each node/VNF and bandwidth requirments of each edge/flow. We perform diffusion on the state sequence based on the state processing described in~\cite{ajay2022conditional}, i.e.,
\begin{equation}
\begin{aligned}
x_{k}(\tau_{t}) = (s_{t},s_{t+1},\dots,s_{t+H-1})_{k}
\end{aligned}
\end{equation}
where $k$ represents the time step in the forward process, and $t$ denotes the time of visiting a state in the trajectory $\tau$. Moreover, we regard $x_{k}(\tau)$ as the noise state sequence from a trajectory of length $H$. We represent $x_{k}(\tau)$ as a two-dimensional array, with each column corresponding to a time step in the sequence. Figure~\ref{fig:example2} visualizes this state sequence representation of the optimal SFC placing-scheduling solution shown previously in Figure~\ref{fig:example2}. 

However, in deploying SFC, it is necessary to extract the properties of the graph for $s_{t}$ and maintain the graph's connectivity. Therefore, we consider both the graph's inherent properties and the SFC's properties.
\begin{equation}
\begin{aligned}
s_{t} = [G_{t};F_{t}]
\end{aligned}
\end{equation}
where $G_{t}$ is an $n+n^{2}$ vector representing the information of the entire graph at time $t$, including graph connectivity and resources on nodes and edges. $F_{t}$ is a vector containing the information of the SFC at time $t$, including the start node of each SFC and the arrival time.
At the same time, according to ~\cite{vignac2022digress}, we further decompose the state as follows:
\begin{equation}
\begin{aligned}
s_{t} = [V_{t};E_{t};F_{t}]
\end{aligned}
\end{equation}

For $F$, since there is not much interrelation among the elements of $F$ and $F$ is more like image-type information, we use traditional Gaussian noise.
However, our model handles graphs with node attributes and edge attributes for $V_{t}$ and $E_{t}$. Therefore, we use a discrete diffusion model instead of an image diffusion model.
Similar to image diffusion models where noise is applied independently on each pixel, we diffuse on each node and edge feature separately. For any node (and similarly for edges), the transition probabilities are defined by the matrices $ [Q_k^V]_{ij} = q(v_k = j \mid v_{k-1} = i) $ and $[Q_k^E]_{ij} = q(e_k = j \mid e_{k-1} = i) $. Adding noise to form $ G_k = (V_k, E_k) $ simply means sampling each node and edge type from the categorical distributions defined by:

\begin{equation}
\begin{aligned}
q(G_{k}|G_{k-1}) = (V_{k-1}Q^{V}_{k},E_{k-1}Q_{k-1}^{E})
\end{aligned}
\end{equation}
This method of noise propagation can effectively diffuse the connectivity information in the graph structure to other nodes.

\vspace{0.05in}
\begin{Lemma} Permutation Invariance~\cite{vignac2022digress}. Our network graph representation satisfies permutation invariance, i.e., reordering of the nodes can lead to the same graph representation.
\end{Lemma}
\vspace{0.05in}

To validate this permutation invariance, the steps are similar to those in~\cite{vignac2022digress} and omitted here due to space limit. The property implies that by reordering the nodes in the network, different matrices can represent the same graph. Learning diffusion models on such data can be very efficient, since it does not require augmenting the data with permutations and gradient updates remain unchanged.

For obtaining the action, we can use the inverse dynamics model ~\cite{agrawal2016learning} to estimate the action from the continuous state at any time step $t$ in the estimated $x_{0}(\tau)$ of the above diffusion model. This can be represented as represented by:
\begin{equation}
\begin{aligned}
a_{t} = f_{\phi}(s_{t},s_{t+1})
\end{aligned}
\end{equation}
For our SFC optimization problem, we design the action as an $m\times n$ vector, where $m$ is the number of SFCs processed at once, and $n$ is the number of nodes. 
This is a one-hot vector indicating which nodes the $i$-th SFC will attempt to deploy on. The environment will try to deploy the SFC based on the provided nodes.
 The data used for training $p_{\theta}$ can also be used to train $f_{\phi}$. Thus, the SFC placement and scheduling variables can be readily extracted from the state sequence.

\subsubsection{Classifier-free guidance}
Given that diffusion models represent different trajectories in the dataset, we will discuss how to use them for planning.
To use the model for planning, it is necessary to additionally adjust the diffusion process based on the feature $y(\tau)$

Here, we use a conditional diffusion model conditioned on the return $y(\tau)$ from the discrete dataset. This approach helps to avoid data contamination that might occur from generating $y(\tau)$ through other means.
However, it is inevitable that our dataset might not always contain optimal solutions. If we use imitation learning, it will inevitably lead to contamination of the conditional diffusion model.
To solve this problem, classifier-free guidance~\cite{ho2022classifier} and low-temperature sampling were used. This approach allows us to extract high-likelihood trajectories from the dataset, representing optimal trajectory behaviors rather than the optimal trajectories themselves.

To implement the above method, we typically need to start sampling from Gaussian noise $x_{K}(\tau)$ and gradually sample down to $x_{0}(\tau)$. The sampling formula for each step is as follows:
\begin{eqnarray}
& \hat{\epsilon_{V}},\hat{\epsilon_{E}},\hat{\epsilon_{F}} & =\epsilon_{\theta}(x_{k}(\tau),k)+w(\epsilon_{\theta}(x_{k}(\tau),y,k) \nonumber \\ 
& &  \ \ \ \ \ \ \ - \epsilon_{\theta}(x_{k}(\tau),k))
\end{eqnarray}
The scalar $w$ is applied to $(\epsilon_{\theta}(x_{k}(\tau),y,k) - \epsilon_{\theta}(x_{k}(\tau),k)$ to enhance and extract the optimal parts of the trajectories $\tau$ that exhibit $y$ in the dataset.
Through the design of the algorithm mentioned above, we obtain the decision-making algorithm~\ref{alg:diffusion}. First, we observe the state in the environment. Next, we perform diffusion conditioned on $y$ and the last observed state to generate sampled states, which are then filled into the horizon. Finally, the corresponding actions are predicted using the inverse dynamics model.

\begin{algorithm}
\caption{SFC Decision Generation using Diffusion}
\label{alg:diffusion}
\begin{algorithmic}[1]
\Require Noise model $\epsilon_{\theta}$, inverse dynamics $f_{\phi}$, guidance scale $w$, history length $C$, condition $y$
\Ensure The array $A$ sorted in non-decreasing order
\State Initialize $h\leftarrow Queue(length = C), t\leftarrow 0$
\While{not done}
    \State Observe environment get $G_{t} = (V_{t},E_{t})$ and Other information $F$
    \State Combine $G$ and $F$ into state $s_{t}$; 
    \State $h.insert(s_{t})$;
    \State Sample $G_{K}\sim q_{V}(n)\times q_{E}(n)$
    \State Initialize $x_{F,K}(\tau)\sim \mathcal{N}(0,\alpha I)$
    \For{$k = K \dots 1$}
        \State $x_{k}(\tau)[:length(h)]\leftarrow h$;
        \State $\hat{\epsilon_{V}},\hat{\epsilon_{E}},\hat{\epsilon_{F}}\leftarrow\epsilon_{\theta}(x_{k}(\tau),k)+w(\epsilon_{\theta}(x_{k}(\tau),y,k) - \epsilon_{\theta}(x_{k}(\tau),k))$
        \State $p_{\theta}^{V} \leftarrow Denoise(x_{k}(\tau),\hat{\epsilon_{V}})$
        \State $p_{\theta}^{E} \leftarrow Denoise(x_{k}(\tau),\hat{\epsilon_{E}})$
        \State $(\mu_{F,k-1},\Sigma_{F,k-1}) \leftarrow Denoise(x_{k}(\tau),\hat{\epsilon_{F}})$
        \State $G_{k-1} \sim\prod_{i} p_{\theta}^{V}(v_{i}|G_{k}) \prod_{i,j} p_{\theta}^{E}(e_{ij}|G_{k})$
        \State $x_{F,k-1}\sim \mathcal{N}(\mu_{F,k-1},\alpha\Sigma_{F,k-1})$
        \State $x_{k-1} = [G_{k-1};x_{F,k-1}]$
    \EndFor
    \State Extract $(s_{t},s_{t+1})$ from $x_{0}(\tau)$
    \State Execute $a_{t} = f_{\phi}(s_{t},s_{t+1});t\leftarrow t+1$
\EndWhile
\end{algorithmic}
\end{algorithm}

\subsubsection{Generate different behaviors}
We aim to guide the diffusion process such that $y(\tau)$ leads to a state sequence that satisfies relevant constraints or exhibits specific behaviors. First and foremost, we want our decisions to achieve the maximum returns. Secondly, we want the diffused states to meet the constraints. Therefore, based on these two requirements, we design the desired $y(\tau)$.
\textbf{Maximizing Returns} To generate trajectories that maximize return, we condition the noise model on the return of a trajectory so $\epsilon_{\theta}(x_{k}(\tau),y(\tau),k) = \epsilon_{\theta}(x_{k}(\tau),R(\tau),k)$,
Where $R(\tau)$ is the return obtained by the trajectory $\tau$.
\textbf{Satisfying Constraints} 
Trajectories may satisfy a variety of constraints, each represented by the set $\mathcal{C}_{i}$, such as reaching a specific goal, visiting states in a particular order, or avoiding parts of the state space.
To generate trajectories satisfying a given constraint $\mathcal{C}_{i}$, we condition the noise model on a one-hot encoding so that $\epsilon_{\theta}(x_{k}(\tau),y(\tau),k) = \epsilon_{\theta}(x_{k}(\tau),{\bf 1}(\tau \in \mathcal{C}_{i}),k)$

Based on the two aforementioned schemes, we combined them to obtain $y(\tau)$ that can guide the generation of the maximum return while satisfying the constraints. The formula is as follows:
\begin{equation}
\begin{aligned}
\epsilon_{\theta}(x_{k}(\tau),y(\tau),k) = \epsilon_{\theta}(x_{k}(\tau),R(\tau)\cdot{\bf 1}(\tau \in \mathcal{C}_{i}),k)
\end{aligned}
\end{equation}

\subsubsection{Training the network diffuser:}
The network diffuser is a conditional generative model we use for decision-making, trained in a supervised manner.
Given a trajectory dataset $\mathcal{D}$, each trajectory is labeled with its achieved return, satisfied constraints, or demonstrated skills.
We simultaneously train the reverse diffusion process $p_{\theta}$ through the noise model $\epsilon_{\theta}$, both of which are divided into three parts based on the definition of the state, and the inverse dynamics model $f_{\phi}$ is parameterized with the following loss:
\begin{equation}
\begin{aligned}
&\mathcal{L}(\theta,\phi)\\ 
=&\mathbb{E}[cross-entropy(v,p_{\theta}^{V})] + \lambda\mathbb{E}[cross-entropy(e,p_{\theta}^{E})]\\
+&\mathbb{E}_{k,\tau\in \mathcal{D},\beta}||\epsilon_{F} - \epsilon_{F,\theta}({\small x_{k}(\tau),(1-\beta)y(\tau)+\beta \varnothing,k})||^{2}\\
+& \mathbb{E}_{(s,a,s^{'})\in \mathcal{D}}||a-f_{\phi}(s,s^{'})||^{2}
\end{aligned}
\end{equation}
where we use $\beta\sim Bern(p)$. For each trajectory $\tau$, we first sample noise $\epsilon\sim \mathcal{N}(0,I)$ and timestep $k\sim\mathcal{U}\{1,...,K\}$.
Then, we construct a noisy state array $x_{k}(\tau)$, and finally predict the noise as $\hat{\epsilon}_{\theta} = \epsilon_{\theta}(x_{k}(\tau),y_{\tau},k)$
Note that in Algorithm~\ref{alg:diffusion}, we use denoising, which is the denoising step of low-temperature sampling. We calculate $\mu_{k-1}$ and $\Sigma_{k-1}$ based on the noisy state sequence and the predicted noise.
And use $\alpha \in [0,1)$ to scale the variance, which enables higher quality sampling for low-temperature samples.

\subsection{Inverse Demonstration for Generating Expert Data}

To obtain demonstration with valid problem-solution pairs of the SFC optimization for training, the key idea of our inverse demonstration is that we do not necessarily need to compute a solution given the NP-hard problem in (\ref{opt1})-(\ref{opt1_6}). Instead, we can reverse the process to start with a set of placement/scheduling decision as input and then determine an SFC optimization problem $(G,F,T)$ with network $G$, SFCs $F$, and deadline $T$, which renders the decision feasible and nearly optimal (i.e., by solving a lexcographic min-max optimization of the completion times). This way, we can obtain sufficient expert demonstrations of the SFC optimization much more easily, for training the proposed network diffuser.

More precisely, we will generate random SFCs and their placement/scheduling decisions, and then update our network and constraints accordingly to obtain an SFC optimization that renders the decisions feasible. Next, given this problem-solution pair as a feasible demonstration, we will formulate and solve a lexicographic min-max problem -- by optimizing the SFC completion times $T_i, \forall i$ (and thus deadline $T$) in a lexicographic manner respect to fixed SFC placement. We show that this lexicographic optimization can be transformed into a separable convex objective~\cite{DBLP:conf/infocom/HuangBWLCT15} and then computed by integer programming over the SFC schedule.

\begin{figure}[th] 
\centering 
\includegraphics[width=0.85\linewidth]{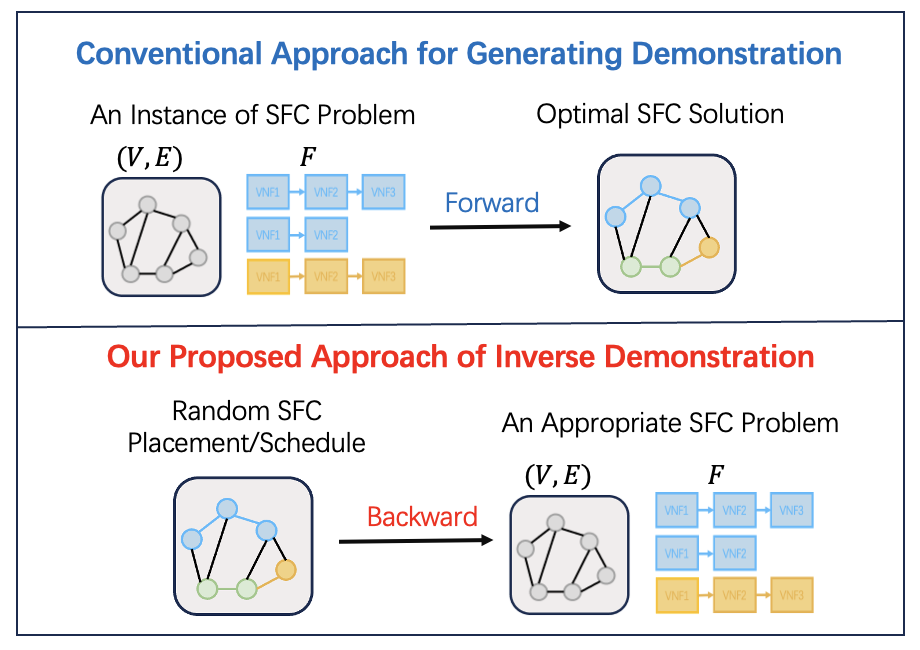} 
\vspace{-0.15in}
\caption{An illustration of our proposed inverse demonstration approach to generate problem-solution pairs for training. Instead of solving a given instance of SFC problem that is NP-hard and has exponential problem space, we start with a randomly generated SFC placement/schedule and then find an appropriate SFC problem that renders the placement/schedule a feasible solution. Further optimization of the demonstration via an integer programming with separable convex objectives yields expert demosntrations for training our network diffuser.} 
\label{fig:algorithm2} 
\end{figure}

In the above algorithm, we first randomly generate an SFC in each step $i$.
Then, we enumerate the nodes in the SFC and use a depth-first search to find a feasible path to deploy them in the network. Next, we update the network (i.e., server resource and bandwidth constraints according to (\ref{opt1_1}) and (\ref{opt1_2})) based on these decisions and collect the sequence of relevant actions and network states as a feasible demonstration trajectory. To further optimize our demonstration, we fixed the SFC placement decisions $z_{i,j}^p$ and consider the following lexicographic min-max optimization of SFC completion times, over the SFC scheduling variables $x_{i,t}$:
\begin{eqnarray}
    & {\rm lex \min } & \max_i (T_i) \nonumber \\
    & {\rm s.t.} & T_i= \max\left\{t\big| x_{i,t} >0, \ \forall t \right\} , \forall i \nonumber  \\
    & & {\rm Constraints \ (\ref{opt1_1}), \ (\ref{opt1_2}), \ (\ref{opt1_3}), \ (\ref{opt1_6}).} \nonumber 
\end{eqnarray}
where $T_i= \max\left\{t\big| x_{i,t} >0, \ \forall t \right\}$ is the completion time of SFC-$i$. Solving this optimization allows us to improve our demonstration with respect to the lexicographic optimization of the SFC completion times. In other words, it 
sequentially minimizes the next-worst SFC completion time as long as it does not affect the
previous completion times.

\begin{figure*}[th]
  \centering
  \begin{minipage}[t]{0.32\textwidth}
      \centering
      \includegraphics[width=\textwidth]{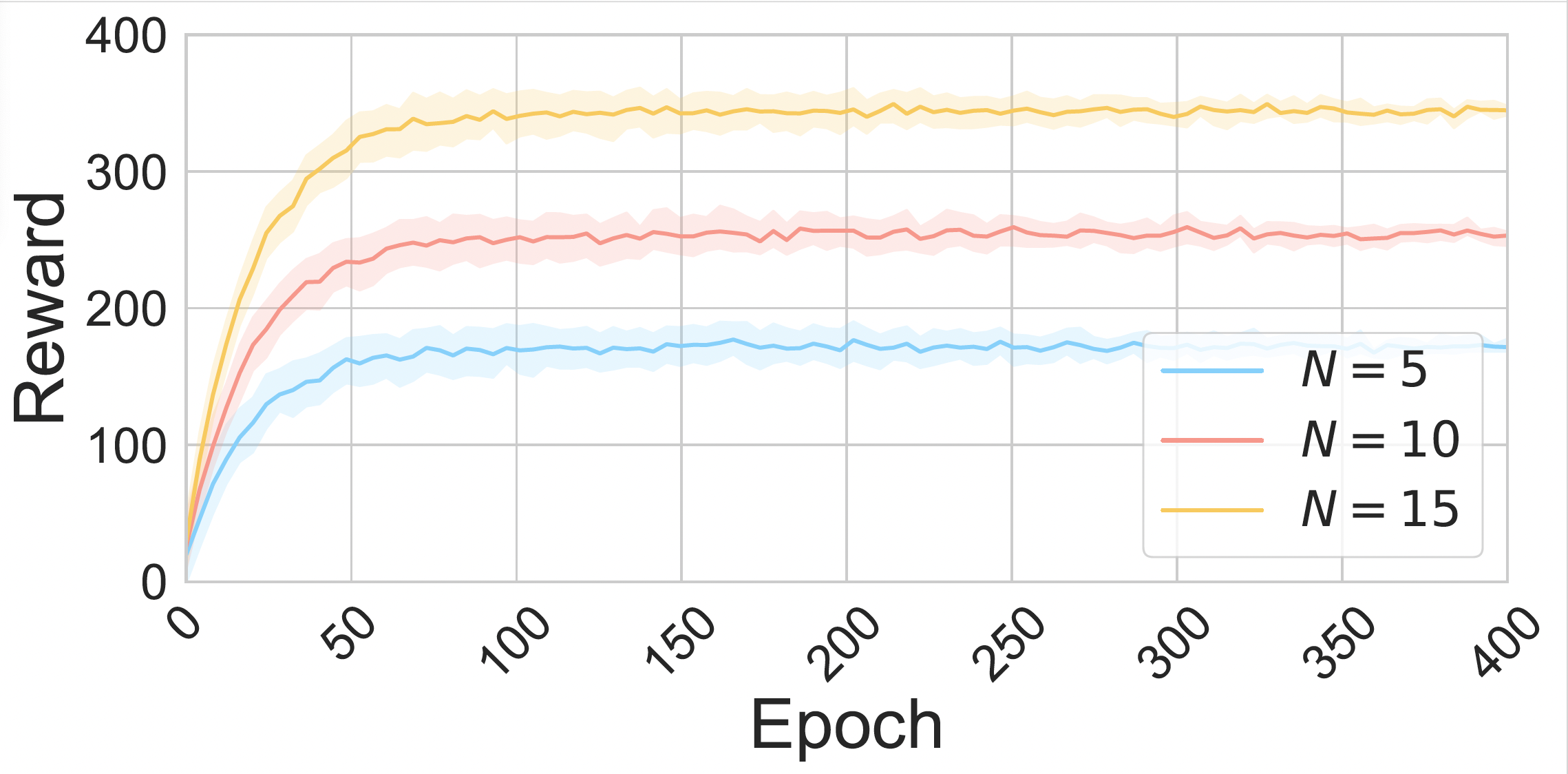}
      \vspace{-0.3in}
     \caption{Achieved reward during training. The SFC placing-scheduling reward (i.e., $\sum_i I_i$) converges within about 100 epochs of training. The reward variance tends to be small and diminishes around 350 epochs, as network size grows from 5 to 15 nodes, showing stable performance. }
      \label{fig:reward_c}
  \end{minipage}
  \hspace{2mm}
  \begin{minipage}[t]{0.32\textwidth}
      \centering
      \includegraphics[width=\textwidth]{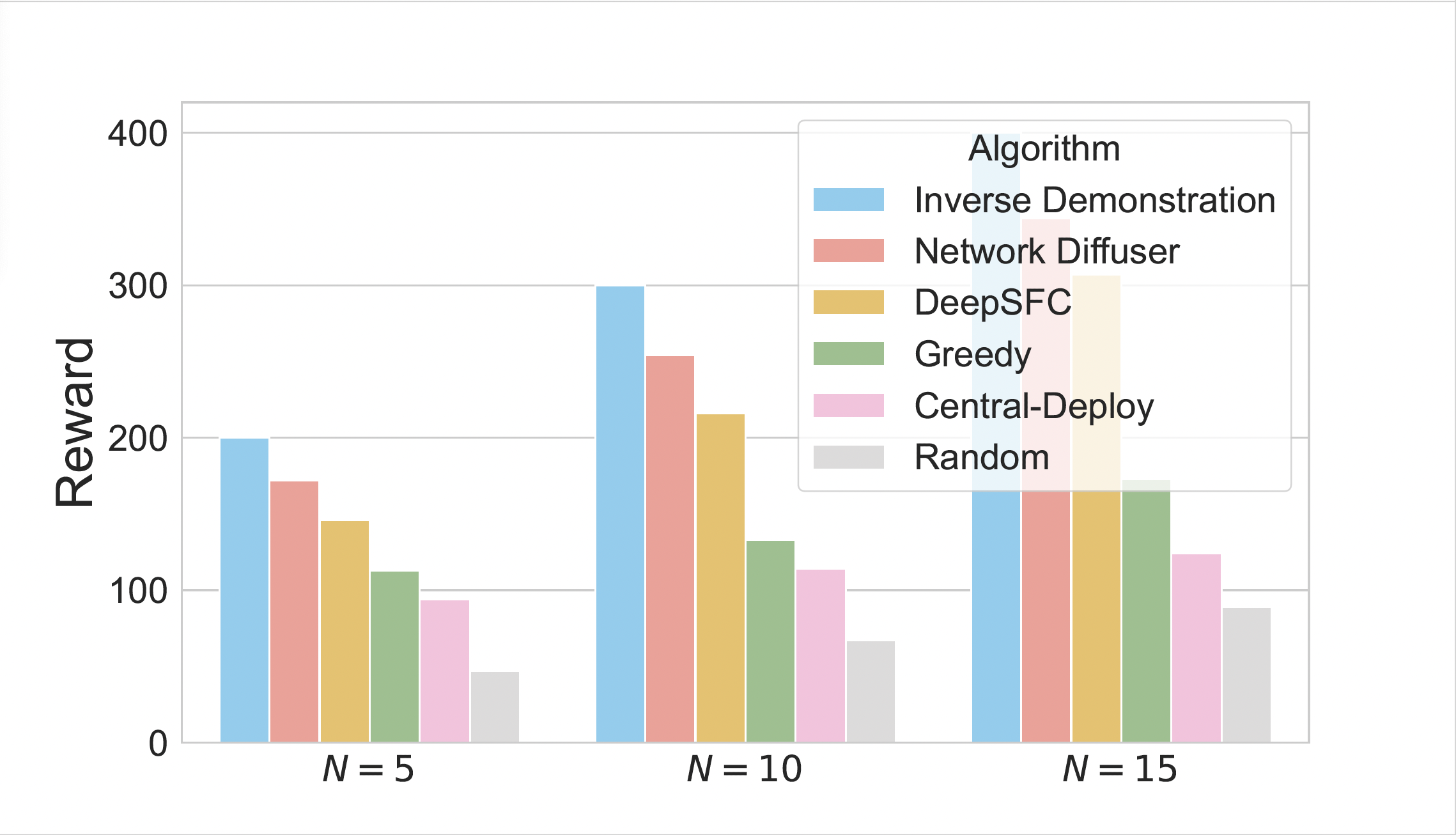}
      \vspace{-0.3in}
     \caption{Comparing network diffuser with heuristic and deep learning baselines. Network diffuser achieves significant reward improvement for varying network sizes. Further, the benefits (i.e., 10\%-15\%) of our inverse demonstration for generating expert training data is validated.}
      \label{fig:reward}
  \end{minipage}
    \hspace{2mm}
  \begin{minipage}[t]{0.31\textwidth}
    \centering
    \includegraphics[width=\textwidth]{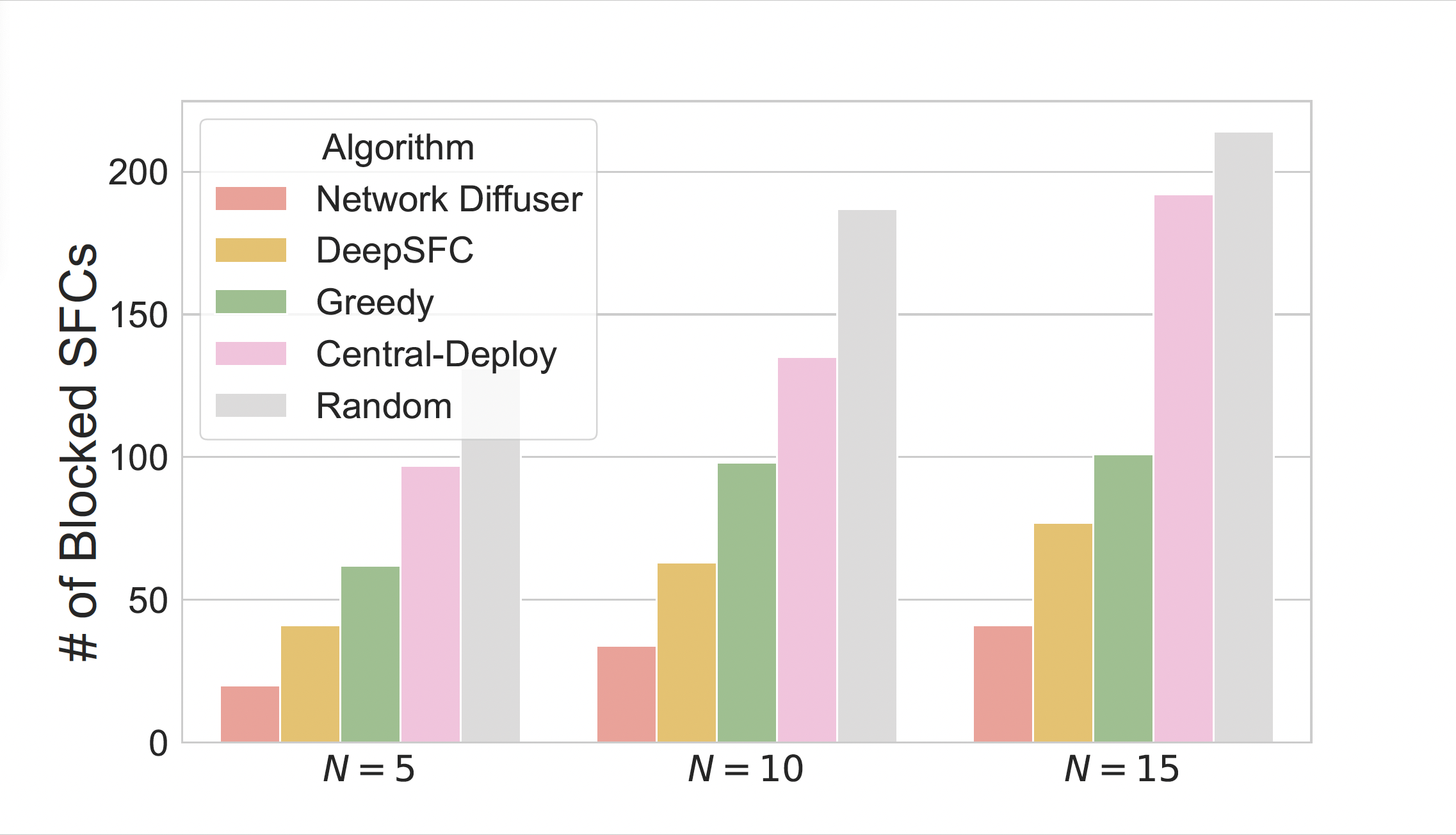}
    \vspace{-0.3in}
    \caption{The number of SFCs blocked in execution. By solving joint placing-scheduling, our proposed network diffuser can significantly reduce the number of SFCs blocked by over 50\% compared to the deep learning baseline. The gap further widens as network size increases. }
    \label{fig:blocked}
  \end{minipage}
  \vspace{-0.1in}
\end{figure*}

We leverage the structure of this lexicographic min-max problem to transform it into an integer programming with a separable convex objective. 
We first eliminate the non-linear completion time constraints $T_i= \max\left\{t\big| x_{i,t} >0, \ \forall t \right\}$ by considering the minimum of $t\cdot x_{i,t} $ over all $t$. It is easy to see that since $x_{i,t}\in \{0,1\}$, we have $T_i=\min_t t\cdot x_{i,t}$. We obtain an equivalent lexicographic min-max problem as:
\begin{eqnarray}
    & {\rm lex \min } & \max_{i,t} (t\cdot x_{i,t}) \nonumber \\
    & {\rm s.t.} & {\rm Constraints \ (\ref{opt1_1}), \ (\ref{opt1_2}), \ (\ref{opt1_3}), \ (\ref{opt1_6}).} \nonumber 
\end{eqnarray}

Next, we consider the calculated $t\cdot x_{i,t}$ terms as a vector $X$ and mathematically define the lexicographic order of two vectors in $\{0,1\}^k$, where $k=NT$ is the dimension of $X$. Let $\vec{X}$ be the vector of $X$ with its elements sorted in non-increasing order. We have that $X$ is lexicographically greater than $Y$ (or equivalently ${X} \succeq {Y}$) if the first non-zero element of $\vec{X}-\vec{Y}$ is positive. Since the objective function of the lexicographic optimization is equivalent to ${\rm lex} \min ( \max (X))$, we can replace it by any function that preserves the lexicographic order. To this end, we consider the following function:
\vspace{0.05in}
\begin{Lemma}
For function $h_X=\sum_{i,t} k^{tx_{i,t}}$, we have $h_{X}\ge h_Y$ if and only if ${X} \succeq Y $.
\end{Lemma}
\vspace{0.05in}
This lemma is straightforward since $X\in \{0,1\}^k$ has dimension $k$ and its elements are either zero or one. Using this lemma we can rewrite the lexicographic optimization as:
\begin{eqnarray}
    & {\min } & \sum_{i}\sum_{t} k^{tx_{i,t}} \nonumber \\
    & {\rm s.t.} & {\rm Constraints \ (\ref{opt1_1}), \ (\ref{opt1_2}), \ (\ref{opt1_3}), \ (\ref{opt1_6})} \nonumber 
\end{eqnarray}
which is an integer programming with a separable convex objective $\sum_{i,t}  k^{tx_{i,t}}$ and linear constraints (\ref{opt1_1}), (\ref{opt1_2}), and (\ref{opt1_3}) over integer variables $x_{i,t}\in \{0,1\}$. This can be computed using off-the-shelf solvers or through relaxation techniques. The result will provide us with an improved demonstration with optimized completion time $T_i$ (and deadline $T$) for the SFC optimization problem.

The pseudo-code for our proposed inverse demonstration approach is shown in Algorithm~\ref{algo1}. We note that this process 
can be repeated multiple times to obtain more complex and superior demonstrations. 

\begin{algorithm}
\caption{Inverse Demonstration for SFC Optimization}
\begin{algorithmic}[1]
\Ensure Demonstrations and corresponding SFC optimization
\For{ $i = 1$ to $N$}
        \State Generate a random SFC-$i$.
        \For{each node $\in$ SFC-$i$}
            \State Deploy node $i$ via depth-first search.
            \State Update the network using (\ref{opt1_1}) and (\ref{opt1_2}).
            \State Update reward $\sum_i I_i$.
            \State Calculate completion time $T_i$.
            \State Collect the actions and the states.
        \EndFor
\EndFor
\State Solve ${\rm lex \min \max} \ T_1,\ldots,T_N$.
\State Update optimal SFC schedules.
\State Update SFC optimization with deadline $T=\min_i T_i$.
\end{algorithmic}
\label{algo1}
\end{algorithm}

%% file: 06simulation.tex
\section{Numerical Evaluation}

\input{00table}

We evaluate our proposed network diffuser on SFC placing-scheduling problems in a simulated environment, for varying network size and dynamic SFC arrivals based on a Poisson process. We illustrate the diffusion and decsion extraction process and compare our network diffuser with a number of baselines using heuristics and deep learning like~\cite{yuan2023deploy}. We note that we do not consider methods like~\cite{DBLP:conf/infocom/TomassilliGHP18,DBLP:conf/infocom/MaoSY22a}, which either consider an offline SFC problem or only focus on SFC placement alone. Our proposed network diffuser significantly outperforms baselines in terms of reward, request waiting time, and non-blocking rate.

\vspace{0.05in}
{\noindent \bf Baselines:} Our evaluation will consider a number of baselines: (i) \textit{DeepSFC:} This approach uses reinforcement learning to solve the SFC deployment problem. For more details, refer to~\cite{yuan2023deploy}.
(ii) \textit{Greedy:} In a resource-constrained edge environment, a greedy strategy is used to find the deployment of the SFC. It starts the DFS from the node with the most remaining resources and selects the node with the most remaining resources at each step.
(iii) \textit{Central-Deploy:} Unlike the greedy strategy, our SFC deployment starts from the central node as much as possible and continues the deployment along the central nodes.
(iv) \textit{Random:} Based on the deployable set of VNFs and considering resource constraints, VNFs are randomly deployed on edge servers.

\vspace{0.05in}
{\noindent \bf Environment:}
All experiments are conducted on a Linux machine with AMD EPYC 7513 32-Core Processor CPU and an NVIDIA RTX A6000 GPU, implemented in python3 and compiled by a Python compiler.
\vspace{0.05in}

\begin{figure}[th]
\centerline{
\includegraphics[width=\linewidth]{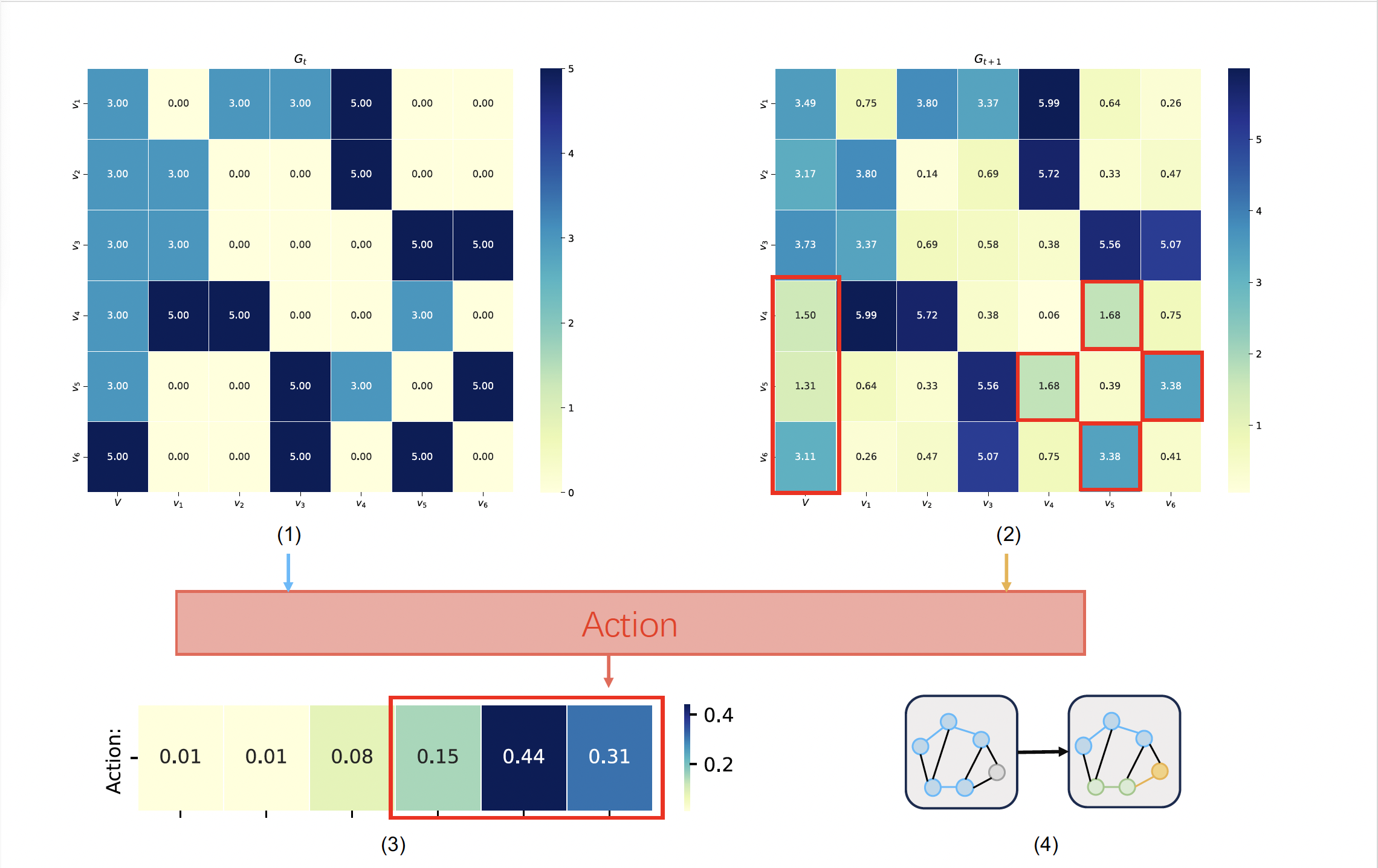}
}
\vspace{-0.4cm}
\caption{
A visualization of generated network state sequence and decision extraction. The heatmap shows remaining capacity on server nodes and links in the network. By contrasting two states at $t$ and $t+1$ in the generated sequence, we can recover SFC placing-scheduling actions from the states.
}
\label{fig:G}
\end{figure}

Figure~\ref{fig:G} shows a visualization of the generate state sequence and decision extraction in our experiments. The heatmap illustrates two future states at time $t$ and $t+1$ respectively. By analyzing the changes in resource utilization in these two states (as highlighted by red boxes), we can determine the SFC placement and schedule decisions. During execution, we employ the trained network diffusion model upon each SFC arrivals, to generate future state sequences and then extract SFC placement and schedule decisions for execution.

We evaluate the convergence of our network diffuser, as the network size grows from $N=5$ to $N=15$ and the number of SFC requests grows from $M=200$ to $M=400$ accordingly. Figure~\ref{fig:reward_c} shows that the SFC placing-scheduling reward (i.e., $\sum_i I_i$) converges within about 100 epochs of training in all three cases. Further, the reward variance (represented by the shaded area) tends to be small and diminishes around 350 epochs, showing stable performance. We also plot the convergence of diffusion training loss $\mathcal{L}(\theta,\phi)$. Again, for different network size, the training loss of our network diffuser using inverse demonstrations converges in a few hundred epochs, demonstrating stable performance.

\begin{figure}[th]
\centerline{
\includegraphics[width=0.8\linewidth]{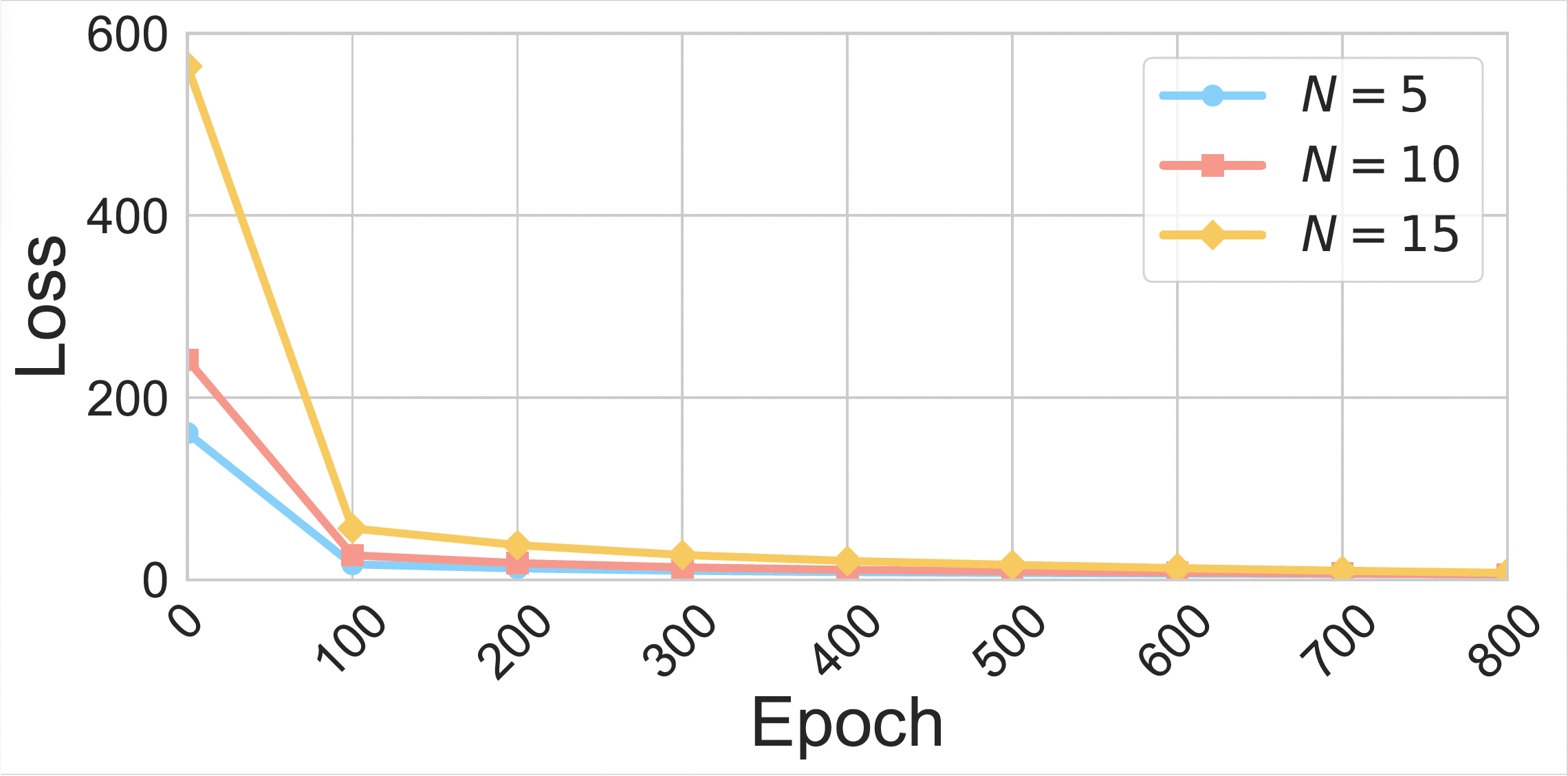}
}
\vspace{-0.3cm}
\caption{Convergence of diffusion training loss $\mathcal{L}(\theta,\phi)$. For different network size, the training loss of our network diffuser using inverse demonstrations converges in a few hundred epochs, demonstrating stable performance.
}
\label{fig:converge}
\end{figure}

We compare our network diffuser with the heuristic and learning baselines, for randomly generated SFC placing-scheduling problems with $T=1000$, as the network size grows from $N=5$ to $N=15$ and the number of SFC requests grows from  $M=200$ to $M=400$. Since the SFC optimization is NP-hard and has a problem space of $|V|^{LM}\cdot T^M$, these are indeed very challenging optimization problems. Figure~\ref{fig:reward} shows that network diffuser achieves significant reward improvement for varying network sizes, with almost 20\% increases in reward. Further, we also conduct ablation study to show the benefits of our inverse demonstration for generating expert training data. It is shown in Figure~\ref{fig:reward} that using training data from optimized inverse demonstration (in Section~IV.B) leads to 10\%-15\% improvement over a scheme that leverages only heuristic-generated data. Thus, the benefits of our inverse demonstration is validated.

Finally, Figure~\ref{fig:blocked} shows the number of SFCs blocked during the execution of different algorithms. By solving the joint placing-scheduling optimization, our proposed network diffuser is able to significantly reduce the number of SFCs blocked by over 50\% compared to the deep learning baseline. The gap further widens as network size increases. The comparison results are summarized in Table~\ref{tab:my-table}. In terms of reward, SFC waiting time, the number of blocked SFCs, and placing-scheduling success rate, our network diffuser demonstrates significant improvement in all aspects for varying network size. This validates the stable benefits of network diffusion for SFC placing-scheduling optimization.

\section{Conclusions}

This paper proposes a novel network diffuser.
It formulates the SFC placing-scheduling optimization as a problem of generating a state sequence for planning. By incorporating SFC optimization constraints and objectives as conditions, we perform graph diffusion on the state trajectories, from which the SFC decisions are then extracted. Further, to address the lack of demonstration data due to NP-hardness and exponential problem space, we develop a novel inverse demonstration approach, by starting with randomly-generated solutions as input and then determining appropriate SFC optimization problems that render these solutions feasible and can be further optimized using integer program with seperable convex objective functions. Significant performance improvements are observed in our numerical experiments.
Both the proposed network diffuser and inverse demonstration methods have the potential to be applicable to a wide range of network optimization problems.

%% file: 00table.tex
\begin{table*}[h!]
\label{tab:smac}
\centering
\resizebox{0.75\textwidth}{!}{%
\begin{tabular}{@{}cclcccc@{}}
\toprule\toprule
Environment & Algorithm    &  & Reward& Average waiting time & Blocked SFCs & Efficiency  \\     \midrule
$N=5$      & Network Diffuser &  & {\bf 172}  & {\bf 1.04}  & {\bf 20} & {\bf 81.7\% } \\ 
$N=5$      & DeepSFC &  & 146  & 2.36  & 41 & 71.1\%  \\ 
$N=5$     & Greedy         &  & 113  & 3.55  & 62 & 60.3\% \\ 
$N=5$      & Central-Deploy &  & 94  & 6.05  & 97  & 52.0\% \\ 
$N=5$      & Random &  & 47  & 8.19  & 131   & 32.5\%  \\ 
\midrule
$N=10$       & Network Diffuser &  & {\bf 254}  & {\bf 1.86} & {\bf 34} & {\bf 80.2\%}  \\ 
$N=10$       & DeepSFC &  & 216  & 2.97  & 63  & 69.3\% \\ 
$N=10$        & Greedy &  & 133  & 4.89  & 98  & 56.7\%  \\ 
$N=10$        & Central-Deploy &  & 114  & 5.96  & 135  & 44.8\% \\ 
$N=10$        & Random &  & 67  & 8.42  & 187  & 30.2\% \\ 
\midrule
$N=15$        & Network Diffusser &  & {\bf 344}  & {\bf 2.01}  & {\bf 41}  & {\bf 77.4\%} \\ 
$N=15$        & DeepSFC &  & 307  & 3.13  & 77    & 62.9\%  \\ 
$N=15$        & Greedy &  & 173  & 5.03  & 101     & 54.1\% \\ 
$N=15$        & Central-Deploy &  & 124  & 6.71  & 192 & 44.3\%  \\ 
$N=15$        & Random &  & 89  & 8.96  & 214  & 29.9\% \\ 
\bottomrule\bottomrule
\end{tabular}%
}
\caption{Our network diffuser outperforms heuristic and deep learning baselines, in terms of reward, SFC waiting time, the number of blocked SFCs, and placing-scheduling success rate. Significant improvements are observed for varying network sizes.}
\label{tab:my-table}
\end{table*}